\documentstyle[prb,aps,epsfig]{revtex}

\tighten       

\begin{document}
\title{Raman phonons as a probe of disorder, fluctuations and local structure in
doped and undoped orthorhombic and rhombohedral manganites.}
\author{L. Mart\'{\i}n-Carr\'{o}n, A. de Andr\'{e}s, M.J. Mart\'{\i}nez-Lope, M.T.
Casais and J.A. Alonso}
\date{Received \today }
\address{Instituto de Ciencia de Materiales de Madrid (Consejo Superior de\\
Investigaciones Cient\'{\i}ficas) Cantoblanco, E-28049 Madrid, Spain}
\maketitle

\begin{abstract}
We present a rationalization of the Raman spectra of orthorhombic and
rhombohedral, stoichiometric and doped, manganese perovskites. In particular
we study RMnO$_{3}$ (R= La, Pr, Nd, Tb, Ho, Er, Y and Ca) and the different
phases of Ca or Sr doped RMnO$_{3}$ compounds as well as cation deficient
RMnO$_{3}$. The spectra of manganites can be understood as combinations of
two kinds of spectra corresponding to two structural configurations of MnO$%
_{6}$ octahedra and independently of the average structure obtained by
diffraction techniques. One type of spectra corresponds to the orthorhombic
Pbnm space group for octahedra with cooperative or dynamic Jahn-Teller
distortions, with stretching modes as the main features and whose
frequencies correlate to Mn-O distances. The other spectrum is associated to
regular but tilted octahedra whose modes can be described in the
rhombohedral R$\overline{3}c$ structure, where only bending and tilt modes
are observed. The main peaks of compounds with regular MnO$_{6}$ octahedra,
as CaMnO$_{3}$, highly Ca doped LaMnO$_{3}$ or the metallic phases of Ca or
Sr doped LaMnO$_{3}$, are bending and tilt MnO$_{6}$ octahedra modes which
correlate to R-O(1) bonds and Mn-O-Mn angles respectively. In low and
optimally doped manganites, the intensity and width of the broad bands are
related to the amplitude of the dynamic fluctuations produced by polaron
hopping in the paramagnetic insulating regime. The activation energy, which
is proportional to the polaron binding energy, is the measure of this
amplitude. This study permits to detect and confirm the coexistence, in
several compounds, of a paramagnetic matrix with lattice polaron together
with regions without dynamic or static octahedron distortions, identical to
the ferromagnetic metallic phase. We show that Raman spectroscopy is an
excellent tool to obtain information on the local structure of the different
micro or macro-phases present simultaneously in many manganites.
\end{abstract}



\twocolumn     
\narrowtext

\section{Introduction}

\label{intro} 
Manganese perovskites R$_{1-x}$A$_{x}$MnO$_{3}$ (R= La or rare
earth, A= Ca, Sr, Ba...) have recently attracted much interest because of
their colossal magnetoresistance effect \cite{1} that makes these systems
promising for magnetic sensors and reading heads devices. In spite of the
tremendous amount of published studies on this subject, many experimental
facts are not well understood, or have to be reinterpreted. Theories do not
give yet a quantitative description of, for example, the metal-insulator
phase transition, but converge in the necessity to include electron-lattice
interactions, which are able to localize the carriers into small polarons
through their coupling to lattice distortions, disorder or phonons\cite
{ziese}. There is also large consensus on the importance of intrinsic
inhomogeneities and phase coexistence that, in the optimally doped regime,
can be visualized as polaron clusters in the ferromagnetic matrix and
metallic clusters in the insulating phase\cite{dagotto}.

The phonons involved in the theories for these CMR systems are even
stretching modes, like symmetric stretching (breathing mode), or
antisymmetric stretching mode (Jahn-Teller (JT) mode) \cite{jtph}. Both
normal modes are identical to the lattice distortions achieved by Mn$^{3+}$
ions in order to split the e$_{g}$ electronic level (JT effect). It is
therefore important to identify these phonons and study their behavior
through the different structural phases and doping levels.

An anomalous frequency hardening and narrowing of the line-width of the tilt
mode as the temperature decreases through the magnetic transition has been
reported in La$_{0.67}$Ca$_{0.33}$MnO$_{3}$ compound. This behaviour has been
explained by a double exchange mechanism with electron-phonon coupling \cite
{5}. Liarokapis et al. \cite{6} observed, for x$\geq $0.4, the disappearance
of the high frequency stretching modes in the La$_{1-x}$Ca$_{x}$MnO$_{3}$
series. Granado et al. \cite{granado} report, in La$_{1-x}$Mn$_{1-x}$O$_{3}$ 
The softening of the high frequency phonons across Tc and explained it to
be caused by a strong spin-phonon coupling. On the other hand, the Raman
background can give information of the electronic excitations. The change of
the diffusive electronic Raman scattering in the paramagnetic phase has been
attributed to the change from small to large polaron regimes across the para
to ferromagnetic phase transition \cite{2}$^{,}$\cite{3}.

The knowledge of the lattice vibrations and their correlation to the
different phases, local order and conduction mechanisms is of crucial
importance. A clear demonstration of this relation is the frustration of the
insulator to metallic transition only by the isotope substitution of O$^{16}$
by O$^{18}$ in a La$_{1/3}$Nd$_{1/3}$Ca$_{1/3}$MnO$_{3}$ compound\cite{4}.
But there is still controversy on the interpretation of the Raman spectra
even for stoichiometric RMnO$_{3}$ whose spectrum corresponds well to the
expected normal modes for its orthorhombic structure \cite{7}$^{-}$\cite{RMO}.
The 490 cm$^{-1}$ peak has been assigned to a bending\cite{7} or to an
asymmetric stretching \cite{6}$^{,}$\cite{14}. Amelitchev et al. \cite
{amelichev} studied the dependence of the low frequency tilt mode with the
tolerance factor, in many compounds. They assign the broad peaks (around 490
and 610 cm$^{-1})$ in doped compounds to second order Raman scattering.
Nevertheless, these authors state as remarkable the absence of the strong
phonon lines near 490 and 612cm$^{-1}$ that dominate LaMnO$_{3}$ while the modes
below 350cm$^{-1}$ are preserved. A detailed Raman study of CaMnO$_{3}$ has
been published\cite{ramanca} during the publication process of the present
work. We do not coincide with this assignment of the tilt and R modes
(discussion in section IV-B). Bj\"{o}rnsson et al. \cite{9} observe two
different sets of phonons in La$_{0.8}$Sr$_{0.2}$MnO$_{3}$, as well as the
appearance of new narrow peaks in La$_{0.9}$Sr$_{0.1}$MnO$_{3}$ at low
temperatures. Their explanation for x=0.2 is the existence of pronounced
local orthorhombic distortions, in the rhombohedral structure, that vanish
at low temperature. However, they cannot explain the evolution with
temperature of the Raman spectra in the x=0.1 compound. The Raman spectra of
doped or non-stoichiometric manganites in their different structural and
magnetic phases for all doping levels are not well understood.

In the present work we study orthorhombic stoichiometric RMnO$_{3}$ (R= La,
Pr, Nd, Tb, Ho, Er, Y and Ca) as well as A site doped or cation deficient
orthorhombic or rhombohedral RMnO$_{3}$ and for different doping levels.

\section{Experimental Details}

Polycrystalline RMnO$_{3}$ (R= Pr, Nd, Tb, Ho, Er and Y) and CaMnO$_{3}$
samples were obtained by citrate techniques. For R=Pr and Nd, the precursors
were treated at 1100$^{o}$C in a N$_{2}$ flow for 12 h; annealing treatments
in an inert atmosphere were necessary to avoid the formation of
non-stoichiometric Pr (Nd) MnO$_{3+\delta }$ phases, containing a
significant amount of Mn$^{4+}$. For R=Tb, the precursor powders were heated
at 1000$^{o}$C in air for 12 h. Finally, low temperature treatments were
necessary for the samples with R=Ho, Y and Er, to increase the yield of the
orthorhombic phases, preventing or minimizing the stabilization of
competitive hexagonal RMnO$_{3}$ phases \cite{10}. In the case of
stoichiometric LaMnO$_{3}$ a single crystal was available; it has been
prepared by the floating zone method \cite{11}. The rhombohedral LaMnO$%
_{3+\delta }$ samples were prepared in polycrystalline form by a citrate
technique as described elsewhere\cite{12}, as well as PrMnO$_{3+\delta }$ 
\cite{13}.

Finally, Ca doped LaMnO$_{3}$ compounds were prepared by the classical
ceramic method by heating stoichiometric amounts of La$_{2}$O$_{3}$, MnO$%
_{2} $ and CaCO$_{3}$ for 72 hours at 1400$^{o}$C. Materials so obtained
were quenched in air.

Raman spectra were obtained with a Jobin-Yvon HR 460 monochromator coupled
to a liquid Nitrogen cooled CCD. The excitation light was the 514.5 nm line
of a Spectra Physics Ar-Kr laser. The incident and scattered beams were
focused using an Olympus microscope. A Kaiser SuperNotch filter was used to
suppress the elastic light. The laser power was reduced down to 0.1 mW
(depending on the sample) in order to avoid the local heating of the samples
at the laser spot. We used a continuous flow Oxford Instrument cryostat
CF2102 to perform measurement from 10K to room temperature (RT).

\section{Structure and normal modes}

All the samples studied here present the same Pbnm (D$_{2h}^{16}$, with Z=4)
orthorhombic structure, except La$_{0.7}$Sr$_{0.3}$MnO$_{3}$ and LaMnO$%
_{3+\delta }$ that are rhombohedral with the R$\overline{3}c$ (D$_{3d}^{6}$,
with Z=2) space group\cite{10}$^{,}$\cite{12}$^{,}$\cite{lasr}. The Pbnm
orthorhombic RMnO$_{3}$ compounds are structurally distorted with respect to
the cubic perovskite, in two ways: the MnO$_{6}$ octahedra present a strong
Jahn-Teller cooperative distortion due to Mn$^{3+}$ ions, and the octahedra
are tilted in order to optimize the R-O bond-lengths. CaMnO$_{3}$ has the
same Pbnm structure as the RMnO$_{3}$ series but the Jahn-Teller distortion
is negligible\cite{castr}. As doping RMnO$_{3}$ with Ca or Sr the
Jahn-Teller distortion decreases, as a consequence of the introduction of Mn$%
^{4+}$ cations, and the structure becomes more regular, maintaining the
tilting of the octahedra. For a broad range of Sr doping, or for oxygen
deficient samples, the reported structure is rhombohedral with tilted
octahedra and strictly identical Mn-O bonds\cite{lasr}.

The Raman active modes of the Pbnm structure (the mirror plane 'm' is
perpendicular to the long c-axis) are: 7A$_{g}$+ 7B$_{1g}$+ 5B$_{2g}$+ 5B$%
_{3g}$. Mn ions do not participate to any Raman mode, as they are located at
inversion centers, while La and O(1) ions display the same kind of
movements. These 24 Raman active modes can be classified into two symmetric
and four antisymmetric stretching modes, four bending modes and six rotation
and tilt modes of the octahedra. At last, eight modes are related to A site
movements.

There are 30 normal modes at the zone center for the R$\overline{3}c$ (Z=2)
rhombohedral structure A$_{1g}$+ 3A$_{2g}$+ 2A$_{1u}$ + 4A$_{2u}$+ 4E$_{g}$+
6E$_{u}$. Among these, 1A$_{g}$+ 4E$_{g}$ are Raman active modes, 3A$_{2u}$+
5E$_{u}$ are IR active and the remaining 2A$_{1u}$+ 3A$_{2g}$ are silent
modes. For this structure, the Raman active modes can be classified into 1A$%
_{1g}$+ 1E$_{g}$ rotational or tilt modes, 1E$_{g}$ bending and 1E$_{g}$
anti-stretching of the MnO$_{6}$ octahedra, and the remaining E$_{g}$ is
related to a vibration of A ions\cite{ilievrh}. The symmetric stretching
mode is A$_{2g}$ and therefore not observable.

\section{Results and discussion}

We discuss the assignment of the main Raman peaks of Ca doped and undoped
RMnO$_{3}$ orthorhombic compounds and explain the apparently different
behavior of their frequencies. We have classified the manganese perovskites
depending on the distortion of the Mn-O octahedra. RMnO$_{3}$ compounds
present a strong cooperative JT distortion, a trivalent state for Mn ions,
insulating behavior, high resistivity and are paramagnetic at RT. A second
group corresponds to compounds with a negligible JT distortion either
orthorhombic (as CaMnO$_{3}$, highly doped La $_{1-x}$Ca$_{x}$MnO$_{3}$ x$%
\geq $0.5 or the ferromagnetic metallic phase of x=0.33 compound), or
rhombohedral as La$_{0.67}$Sr$_{0.33}$MnO$_{3}$. At last we discuss the
paramagnetic phase of the 0${<}$x${<}$0.5 compounds as well as non
stoichiometric RMnO$_{3}$ samples and phase coexistence.

\begin{figure}[bp]
\begin{center}
\includegraphics[width=7.5cm]{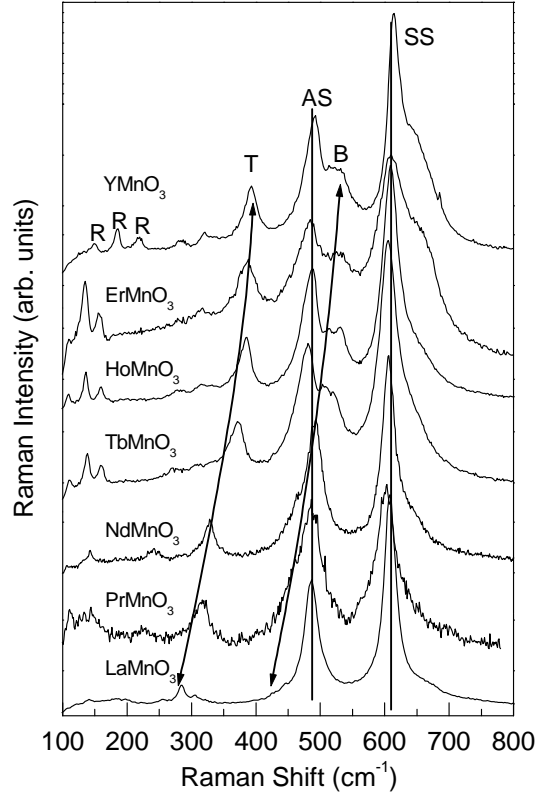}
\end{center}
\caption{Raman spectra at RT of Pbnm RMnO$_{3}$ pellets. LaMnO$_{3}$ is a 
single crystal. Same phonons in the spectra
are connected with vertical lines. R: R ion
modes, T: tilt, B: bending, and AS: antisymmetric and SS: symmetric
stretching.}
\label{Fig1}
\end{figure}

\subsection{Compounds with strong cooperative Jahn-Teller distortion: RMnO$%
_{3}$}

We have analyzed the Raman spectra at room temperature of orthorhombic RMnO$%
_{3}$ compounds as a function of the chemical pressure by changing the rare
earth ion (Fig. 1). There is agreement between all authors in the assignment
of the peak around 610 cm$^{-1}$, related to a symmetric stretching of the
basal oxygen ions of the octahedra (B$_{1g}$), and that around 280 cm$^{-1}$, 
related to some octahedra tilt, but there are several claims concerning
the one around 480 cm$^{-1}$ and some confusion in the bending, tilt and
rock modes in doped compounds\cite{6}$^{,}$\cite{7}$^{,}$\cite{amelichev}. We 
assigned the 480 cm$^{-1}$ peak to an antisymmetric
stretching (A$_{g}$) associated with the JT distortion \cite{RMO}$^{,}$\cite{14} and give
here further evidence. Regarding tilt and bending modes, different
combinations of displacements of the in plane O(2) ions along z axis and of
the apical O(1) oxygen perpendicular to it, give rise to four tilt modes (A$%
_{g}$+ B$_{1g}$) and (B$_{2g}$+ B$_{3g}$) and two bending modes (A$_{g}$+ B$%
_{1g}$). We have grouped in brackets the Raman modes that may have very
similar frequencies because they correspond to the same movements around x
or y-axis. Raman spectra corresponding to modes with B$_{2g}$ and B$_{3g}$
symmetries have a very low intensity \cite{7}, so we will focus on A$_{g}$
and B$_{1g}$ modes. The peaks at the lowest energies (marked with R in
Fig.1) correspond to rare earth movements. Table \ref{tabla1} collects the
measured frequencies together with the values estimated for the three R
modes ($\omega _{R}$=$\omega _{Y}\sqrt{M_{Y}/M_{R}}$) considering only the
effect of the R mass and taking Y compound frequencies as references. The
three R modes behave exactly as expected in all measured compounds The
differences between estimated and measured frequencies are lower than 2 cm$%
^{-1}$ except for LaMnO$_{3}$ ( 6 and 9 cm$^{-1}$), due to the increase of
La-O(1) distances compared to Y-O(1) ones.

\begin{figure}[bp]
\begin{center}
\includegraphics[width=7.5cm]{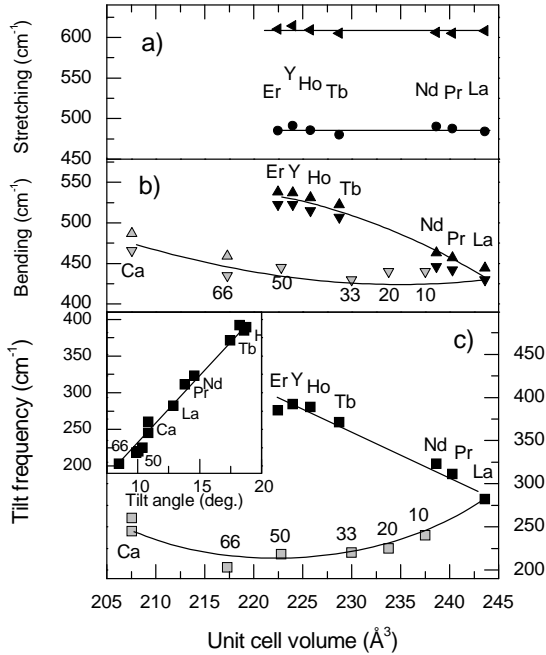}
\end{center}
\caption{Dependencies, vs the unit cell volume, of the RT frequencies of the
symmetric and antisymmetric stretching modes a), bending modes b) and tilt
mode c) for RMnO$_{3}$ (black symbols) and La$_{1-x}$Ca$_{x}$MnO$_{3}$
series (grey symbols). The numbers in the figure indicate the percentage of
Ca doping. The inset shows the linear dependence of the tilt mode frequency
with the tilt angle, which is defined as (180-(Mn-O(1)-Mn))/2 }
\label{Fig2}
\end{figure}

\begin{figure}[bp]
\begin{center}
\includegraphics[width=7.5cm]{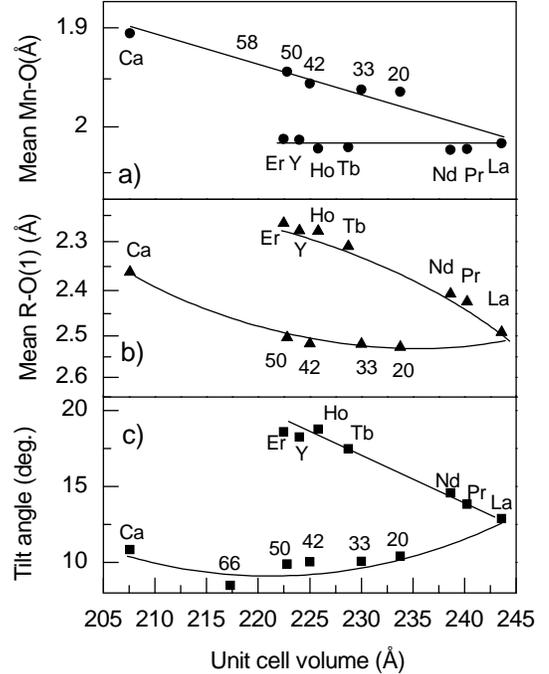}
\end{center}
\caption{Dependencies, vs the unit cell volume, of mean Mn-O distances a),
mean R-O(1) distances b) and the octahedra tilt angle (defined as (180-(Mn-O(1)-Mn))/2) c) for RMnO$_{3}$
(structural data from Refs. 18 and 19) and La$_{1-x}$Ca$_{x}$MnO$_{3} $ series (structural data from Refs. 6 and
26-28). The numbers in the figure indicate the percentage of Ca doping.}
\label{Fig3}
\end{figure}

In Fig.1 the two most intense peaks (610 and 480 cm$^{-1}$) vary only
slightly (less than 10 cm$^{-1}$) with the rare earth (Fig. 2). The symmetric (around 610
cm$^{-1}$) and the antisymmetric, or Jahn-Teller, (around 480 cm$^{-1}$)
stretching modes involve nearly pure Mn-O bond stretching therefore a simple
(dMn-O)$^{-1.5}$ dependence of these phonon frequencies is expected \cite
{deandres} and observed. This is an indication that the MnO$_{6}$ octahedron
volume and bonds are insensitive to quite large changes in the tolerance
factor in perovskites when comparing compounds with the same Mn valence
state (in this case 3$^{+}$). On the contrary, the frequencies of the peaks
labelled with ''B'' and ''T'' increase strongly (over 100cm$^{-1}$) when the cell volume shrinks
(Fig. 2b and c). The behavior of these phonons is parallel to that of the
mean value of the two shortest R-O(1) bonds and to the static octahedra tilt
(deviation from the cubic structure defined as (180-(Mn-O(1)-Mn))/2) respectively (Fig. 3b and c). We assign
the ''B'' peaks (two peaks can be distinguished) to bending modes (A$_{g}$ +B%
$_{1g}$) and the ''T'' peak to (A$_{g}$ +B$_{1g}$) tilt modes. The tilt and
bending modes with A$_{g}$ symmetry, that correspond to rotations around
y-axis, are plotted in Fig. 4a and Fig. 4b, respectively. The B$_{1g}$ modes
correspond to the same atomic movements but around x-axis, therefore A$_{g}$
and B$_{1g}$ modes are expected to be very similar in frequency. The
displacement of O(1) in these modes is, in fact, a stretching of R-O(1)
bonds (Fig. 4c). Each O(1) ion is surrounded by four R ions with two short
and two long distances. Fig. 3b displays the evolution of the mean value of
the two short R-O(1) bonds. Comparing Fig. 2b and Fig. 3b the correlation
between the bending mode frequency and the R-O bond length becomes obvious.

\begin{figure}[bp]
\begin{center}
\includegraphics[width=7cm]{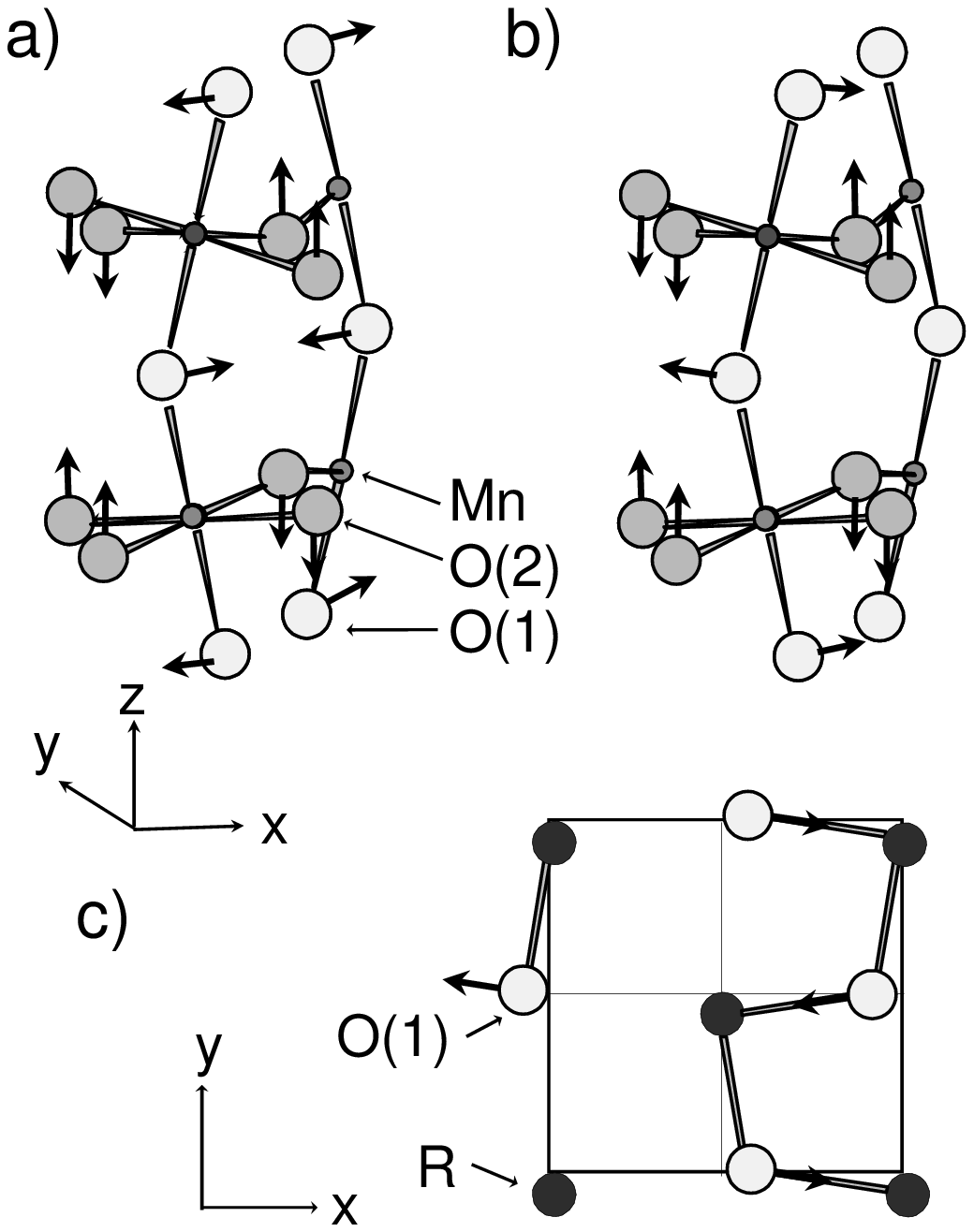}
\end{center}
\caption{ a)A$_{g}$ Octahedra tilt mode, b) and c) A$_{g}$ octahedra bending
mode. Both in the Pbnm structure.}
\label{Fig4}
\end{figure}

In summary, the present data show that stretching mode frequencies correlate
to Mn-O bond distances while the R-O bond and octahedra tilt angle dominate
bending and tilt modes, respectively. These are the expected behaviors for
such normal modes and must be kept in mind when trying to assign the spectra
of doped or related compounds.

\subsection{Insulating paramagnetic La$_{1-x}$Ca$_{x}$MnO$_{3}$ from x=0 to
x=1}

CaMnO$_{3}$ is also described with the Pbnm space group but with important
differences compared to the previous RMnO$_{3}$ compounds, that are the
clues to understand their very different Raman spectra (Fig. 5c, 0\% and
100\% spectra). The Mn valence state is 4+ therefore no static nor dynamic
JT distortions are present since the 3d e$_{g}$ Mn orbital is empty. CaMnO$%
_{3}$ cannot be described with the cubic perovskite structure because the
tilt of the octahedra remains. The two pairs of peaks (around 240 cm$^{-1}$
and 460 cm$^{-1}$) cannot correspond to stretching modes (the dominant ones
in RMnO$_{3}$ ). Taking into account the much shorter Mn-O distances their
frequencies would be around 520 and 660 cm$^{-1}$. We assign these quite
sharp peaks to the bending (A$_{g}$= 487 cm$^{-1}$, B$_{1g}$= 466 cm$^{-1}$)
and tilt (A$_{g}$= 245 cm$^{-1}$, B$_{1g}$= 260 cm$^{-1}$) modes (A$_{g}$
modes are plotted in Fig.4). The frequencies and relevant structural
parameters of Ca doped series are presented in Fig. 2 and Fig. 3 making
obvious their parallel behavior and confirming the modes assignment.

\begin{figure}[bp]
\begin{center}
\includegraphics[width=7.5cm]{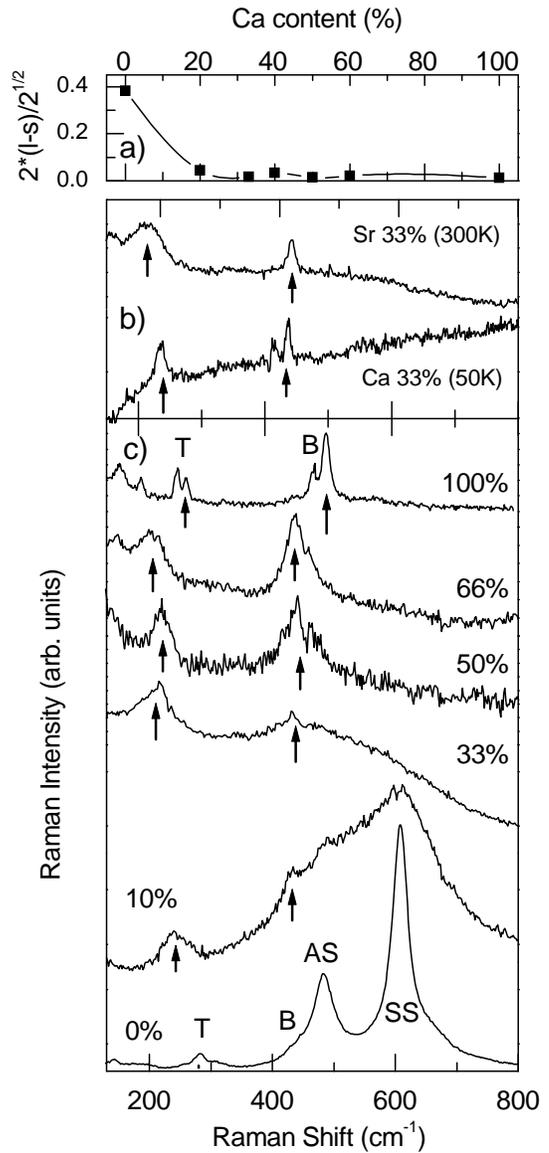}
\end{center}
\caption{ a) Cooperative static JT distortion, from diffraction data in
Refs. 16, 20-22, for La$_{1-x}$Ca$_{x}$MnO$_{3}$ series; b) Raman spectra of
FM metallic Ca (50K) and Sr (300K) doped LaMnO$_{3}$ with x=0.33; c) RT
Raman spectra of La$_{1-x}$Ca$_{x}$MnO$_{3}$ samples, as function of Ca
content. The arrows point the narrow peaks corresponding to the tilt and
bending modes.}
\label{Fig5}
\end{figure}

During the referee process, a study of the Raman spectra of CaMnO$_{3}$
appeared \cite{ramanca}. These authors assigned the doublet at 242 and 258 cm$^{-1}$ 
to A$_{g}$ and B$_{2g}$ modes involving O (1) and Ca vibrations in
the xz plane (in the Pnma representation that correspond to A$_{g}$ and B$%
_{1g}$ modes and xy plane in Pbnm one), and the 184 cm$^{-1}$ peak to the A$%
_{g}$ tilt mode. Looking at Table \ref{tabla1}, the R modes in CaMnO$_{3}$
are expected at 220, 274 and 325 cm$^{-1}$ and, in fact, two peaks of the
correct symmetries are observed at 278\cite{ramanca} and 322cm$^{-1}$ (Fig.
5c and Ref. 16). The Ca-O1 distances are similar to Y-O1 ones, consequently,
the rule that has been found to be valid for all RMnO$_{3}$ series should
be valid for Ca compound. Therefore, the peaks at 242 and 259 cm$^{-1}$,
that show differences compared to the expected values over 22 cm$^{-1}$
(Table \ref{tabla1}), are not R modes. These peaks correspond to oxygen tilt
modes. In a first approximation, as the unit cell volume decreases, the
frequencies of all modes are expected to increase because the bonds shrink.
The reason for the tilt modes to decrease is due to the reduction of the
tilt angle as shown in the inset of Fig.2.

Fig. 5c collects the Raman spectra of LCMO compounds for several doping
levels. Very significant changes are observed. Compared to stoichiometric
LaMnO$_{3}$, for doping levels as low as 10\%, the width of the stretching
modes increases and their spectral weight decreases dramatically with
doping. At 50\% Ca doping and above no stretching modes appear and the
spectra resembles strongly to the CaMnO$_{3}$ one.

\begin{figure}[bp]
\begin{center}
\includegraphics[width=7.5cm]{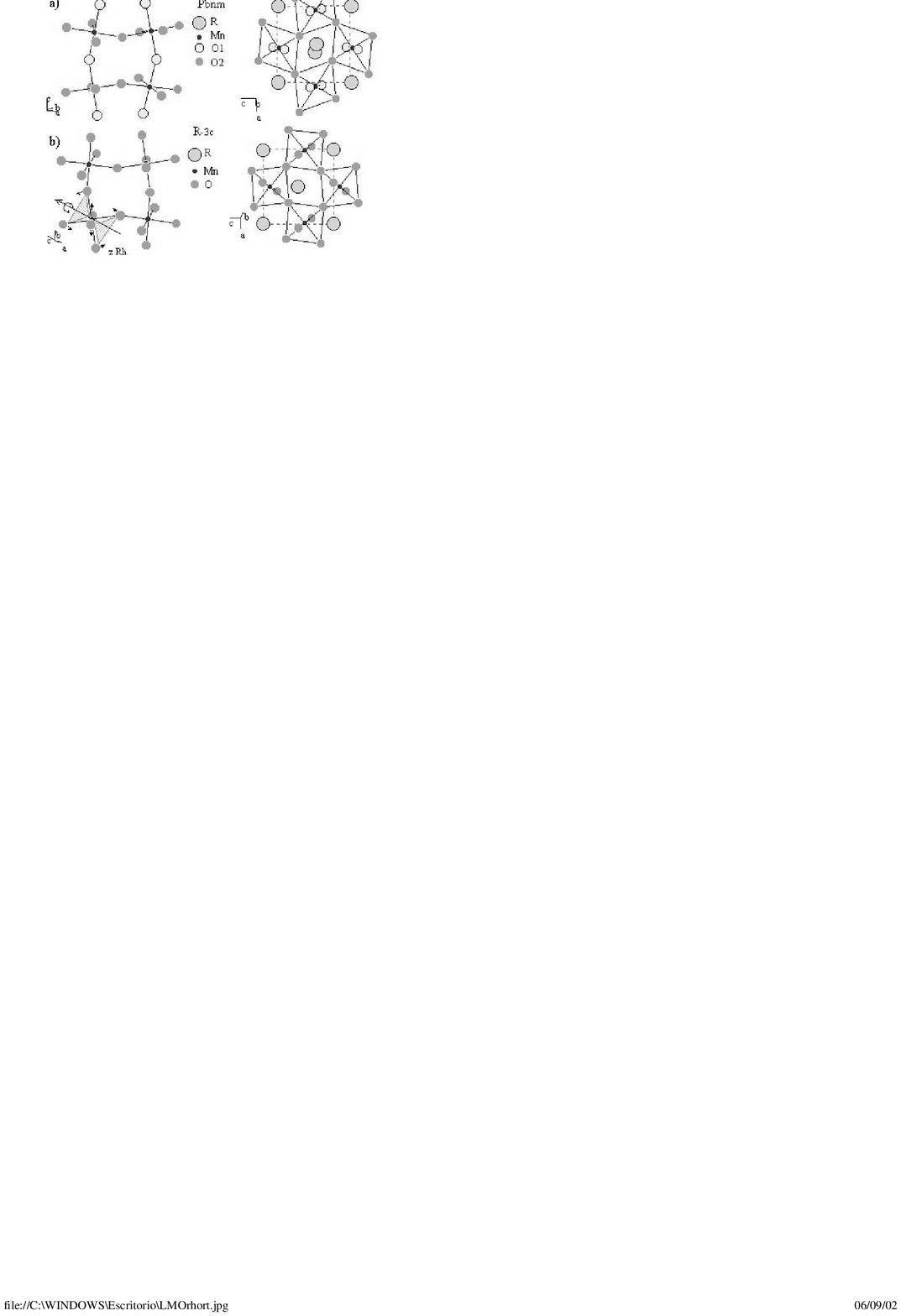}
\end{center}
\caption{ Pbnm orthorhombic (a) and rhombohedral R$\overline{3}c$ (b) RMnO$%
_{3}$ structures. The rhombohedral axes have been rotated in order to better
visualize the similarities between the octahedra network in both structures.
The A$_{1g}$ tilt mode is shown.}
\label{Fig6}
\end{figure}

It is important to note that paramagnetic CaMnO$_{3}$, highly doped
compounds (Fig. 5c) and the ferromagnetic metallic phase of La$_{0.67}$Ca$%
_{0.33}$MnO$_{3}$ (Fig. 5b) present very similar Raman spectra. These quite
different systems have in common that the cooperative JT distortion is
negligible (Fig. 5a) and that they do not present the stretching modes. We
can conclude that a JT distortion, or probably a distortion engendering
significantly different Mn-O bond lengths, is necessary to produce a
measurable intensity of the stretching modes. The allowed modes by global
symmetry considerations are identical to the previous RMnO$_{3}$ compounds
but the change in the polarizability due to a particular mode, that
determines its Raman intensity, is found to vanish for the stretching modes
of the regular octahedra.

On the other hand, these spectra are nearly identical to that corresponding
to rhombohedral La$_{0.7}$Sr$_{0.3}$MnO$_{3}$ (Fig. 5b). The observable
Raman active modes of the R$\overline{3}c$ structure correspond to bending
and tilt modes of the oxygen octahedra, whose atomic displacements are very
similar to the bending and tilt modes of the orthorhombic structure (see
tilt mode in Figs. 4a and 6b). From R$\overline{3}c$ to Pbnm the number of
expected peaks increases. Only one bending E$_{g}$ and two tilt modes (A$_{1g}$ + E$_{g}$) 
are expected in the R$\overline{3}c$ rhombohedral
structure while four bending and four tilts modes correspond to the
orthorhombic Pbnm one. As shown in Fig. 6, the octahedra network in the Pbnm
and R$\overline{3}c$ structures are very similar. Both present tilted
octahedra with similar angles (about 160${^{o}}$ for R$\overline{3}c$ ) but
Pbnm structure can present up to three different Mn-O distances, while all
Mn-O bond lengths are identical in the rhombohedral one. But highly doped Ca
compounds, as well as the metallic phases of Ca and Sr doped compounds, do
not present cooperative or dynamic JT distortions. In these cases the
octahedra in Pbnm and R$\overline{3}c$ space groups are almost identical.
Even if the stretching modes are allowed by symmetry in the Pbnm space
group, the origin of the close similarity of their spectra, in particular
the absence of high frequency modes, becomes clear.

We conclude therefore that there are two spectra type, one corresponding to
compounds with a cooperative JT distortion (LaMnO$_{3}$ type) that is
dominated by the stretching phonons, and the other without these modes
corresponding to compounds without JT distortion, like CaMnO$_{3}$, metallic
Pbnm phases or rhombohedral LSMO.

\subsection{Paramagnetic phases of low to optimally doped manganites}

At low doping levels, for example La$_{0.9}$Ca$_{0.1}$MnO$_{3}$ in Fig 5c or
RMnO$_{3-\delta },$ (R=La or Pr) in Fig. 7, the spectra are similar to LaMnO$%
_{3}$ one but with wider peaks. The dynamic fluctuations of the octahedra
caused by the lattice polaron hopping increase very substantially their
width. But La$_{0.66}$Ca$_{0.33}$MnO$_{3}$ (LCMO) or La$_{0.9}$MnO$_{3}$
(Fig. 7) present some spectral weight in the stretching frequency range (AS and SS)
together with narrower peaks (indicated with arrows). In order to check
wether these peaks have the same origin as the broad high frequency bands,
we compare their spectra to the undoped LaMnO$_{3}$ at T${>}$800K (lower
spectrum in Fig. 7) above the so-called JT transition. At these
temperatures, the cooperative JT distortion is melted but dynamical JT
fluctuations remain \cite{14}$^{,}$\cite{11}. This is a very similar
scenario to what occurs in the paramagnetic phase of LCMO, therefore similar
Raman spectrum would be expected. Fig. 7 shows that above 800 K the
intensity has decreased drastically compared to RT spectrum (Fig. 5) but
some spectral weight is observable in the high frequency region
corresponding to the stretching modes while tilt and bending modes have
almost vanished. It is evident that dynamic distortions of the MnO$_{6}$
octahedra cannot give rise to the quite narrow peaks around 230 and 420 cm$^{-1}$, 
but are the origin of the broad stretching bands.

\begin{figure}[bp]
\begin{center}
\includegraphics[width=7.5cm]{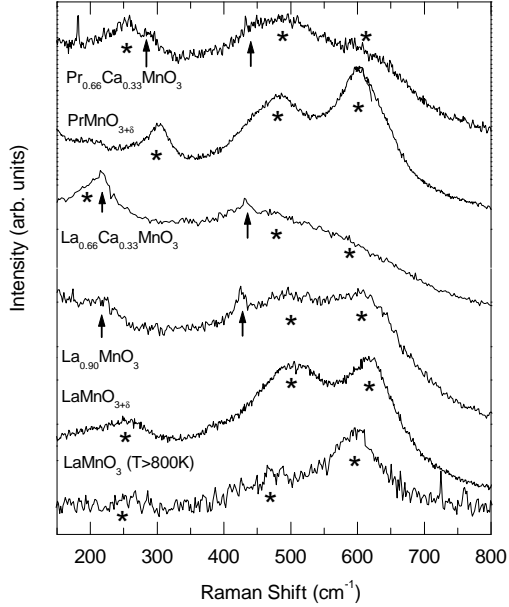}
\end{center}
\caption{ Raman spectra at RT of non-stoichiometric RMnO$_{3}$ (R=La and Pr)
as well as 33\% Ca doped compounds. Bottom: Raman spectrum of LaMnO$_{3}$ at T$>$ 800K above the 
JT transition. Small arrows point the narrow peaks related to the R$%
\overline{3}$c type phase and the stars position the modes associated with
the Pbnm type phase with dynamic and/or cooperative JT distortion.}
\label{Fig7}
\end{figure}

Doping LaMnO$_{3}$ gives rise Mn valence mixing and local static distortions
around the doping ion. Simultaneously, in the paramagnetic phase, the
hopping electrons trapped as lattice polarons induce lattice distortions.
This disorder is of dynamic character and is expected to change at the
insulator to metallic phase transition and has been described as a crossover
from small to large polaron regimes \cite{2}. Diffraction techniques give us
valuable information on the mean interatomic distances but not on
instantaneous or randomly distributed atomic positions, which are important
for the Raman scattering effect. The analysis of the diffuse X-ray
scattering \cite{16} and Reverse Monte-Carlo simulations of neutron
diffraction patterns\cite{louca}$^{,}$\cite{17} have shown that local
octahedron distortions are much larger, around and above Tc, than the
obtained by standard diffraction analysis. Diffraction shows very slight
structural changes at the ferromagnetic metallic transition while Raman
spectra above (Fig. 5c) and bellow (Fig. 5b) Tc are quite different. Above Tc, the
features similar to high temperature LaMnO$_{3}$ correspond to the
paramagnetic matrix dynamically distorted by polaron hopping, and the ones similar to CaMnO$_{3}$ spectrum
indicate that part of the sample has a structure alike 
the metallic ferromagnetic phase. It might correspond to the magnetic
clusters or magnetic polarons that have been observed by different
techniques \cite{19}.The structure of these entities is unknown but present
ferromagnetic correlations. In the present knowledge of phase segregation or
inhomogeneous intrinsic phase, present in most of the manganites, it is
realistic to think that the second spectrum corresponds to ferromagnetic
metallic droplets in the paramagnetic matrix with polaronic (lattice
polarons) conduction. At
low temperatures, well in the metallic regime,
only the narrow peaks, similar to CaMnO$_{3}$, remain. 

\begin{figure}[bp]
\begin{center}
\includegraphics[width=6.5cm]{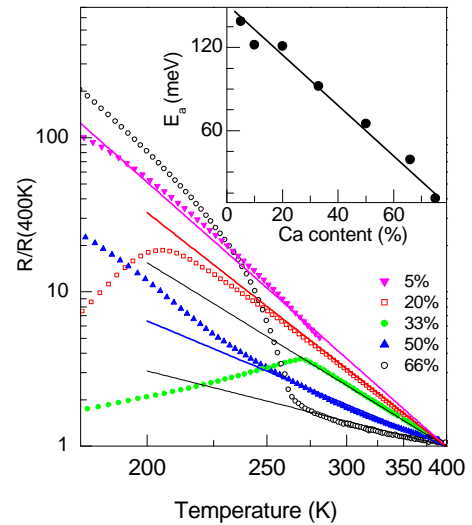}
\end{center}
\caption{ Normalized resistance of samples with several Ca content. Lines
are fits to the paramagnetic state resistance with: R=R$_{0}$exp(E$_{a}$/K$%
_{B}$T). The inset shows the obtained activation energies E$_{a}$ vs Ca
content.}
\label{Fig8}
\end{figure}

The coexistence of the two types of spectra, related to two phases or
micro-phases, is observed in several compounds in Fig. 5c and Fig. 7 and,
for example, in Fig. 1b of Ref. 16. In fact, in the present scenario we can
understand the temperature dependence of spectra from Ref. 17. Compounds
with x=0.2 and 0.1 are both paramagnetic insulators (PMI) at high
temperatures. The x=0.2 one becomes ferromagnetic (FM) and metallic at 280K
while x=0.1 transforms in a FM insulator at 200K. The temperature dependence
of La$_{0.8}$Sr$_{0.2}$MnO$_{3}$ spectrum is ''identical'' to the changes
observed in La$_{0.67}$Ca$_{0.33}$MnO$_{3}$, in spite of their different
average structures (R$\overline{3}c$ and Pbnm, respectively). In the
insulating paramagnetic phases, dynamically distorted octahedra (Pbnm type)
coexist with regular octahedra (R$\overline{3}c$ type) while in the
ferromagnetic metallic regime only regular octahedra remain. On the
contrary, in x=0.1 Sr doped LaMnO$_{3}$, above Tc, only the broad features
are observed corresponding to the Pbnm structure with dynamic distortions
related to polaron hopping. But, below Tc, the appearance of narrows peaks,
in coexistence with the broad bands, indicates that metallic, and probably
ferromagnetic, clusters in the insulating matrix are formed. In the
discussion about the nature of the FMI phase of this x=0.1 compound, on
whether it is a canted ferromagnetic or phase segregation is occurring, Raman
spectrum is consistent with the second hypothesis. Nevertheless a more
detailed analysis is necessary.

When Ca content increases the binding energy of the lattice polarons
decreases, as indicates the reduction of the activation energy for
conduction in the paramagnetic phase (Fig. 8). The reduction of the polaron
energy implies the collapse of the amplitude of the dynamic distortions it
produces in the lattice and, therefore, of the spectral weight of the LaMnO$%
_{3}$ type spectrum. The width of the instantaneous distribution of Mn-O
distances is the origin of the width of the Raman peaks. For doping
concentrations of 50\% and above the lattice is in fact, instantaneously,
much more regular than at low doping. We recover the CaMnO$_{3}$ or LSMO
type spectrum. In the metallic regime, the hopping carriers become
delocalized and do not induce lattice distortions. Sharp phonon peaks are
observed bellow 450 cm$^{-1}$ independently of the crystallographic
structure (compare metallic Pbnm LCMO and metallic R$\overline{3}c$ LSMO at
RT (Fig. 5b) and at low temperature \cite{18}). Therefore, the width and
intensity of the stretching modes are a measure of the amplitude of the
dynamical JT distortions which correlates to the activation energy for
polaronic conduction.

At last, the main characteristic of the Raman spectra of most charge ordered phases, that
is the recovery of the stretching phonon peaks, is understood as being caused
by the orbital order and the concomitant cooperative JT distortion.

\section{conclusions}

The analysis of the Raman spectra of RMnO$_{3}$ Pbnm compounds that present
strong cooperative JT distortion show that the stretching modes (symmetric
and antisymmetric) correlate to Mn-O bonds while bending modes are
determined by the R-O(1) mean distance and tilt modes by Mn-O-Mn angle.
Compounds without cooperative or dynamic distortions of the MnO$_{6}$
octahedra, independently of their mean crystallographic structure, do not
present stretching modes and the characteristics of their spectra can be
approached to the R$\overline{3}c$ space group vibrations. The bending and
tilt normal modes are equivalent to the Pbnm ones and the measured
frequencies follow the same rules as RMnO$_{3}$ ones. This explains the
similarities between insulating CaMnO$_{3}$ (Pbnm) and the metallic phases
of orthorhombic La$_{0.67}$Ca$_{0.33}$MnO$_{3}$ and rhombohedral 
La$_{0.67} $Sr$_{0.33}$MnO$_{3}.$ The activation energy for conduction in the insulating
phases, which is proportional to the polaron binding energy, is a measure of
the amplitude of the dynamic distortion it produces in the lattice.
Therefore, as the Ca content increases, the activation energy decreases as
well as the amplitude of the dynamic distortions. The width of the peaks
corresponding to RMnO$_{3}$ type spectrum increases and their intensity
decreases as the Ca content rises up to about 50\% where these peaks vanish
and only the CaMnO$_{3}$ type spectrum is observed. The identification of
these types of spectra and their correspondence with particular octahedron
configurations are the keys to understand the Raman spectra of most
manganese perovskite. Moreover, this allows to detect the simultaneous
presence of several different phases and to obtain insight in their local
structure.

\begin{acknowledgments}
We wish to acknowledge the financial support from CICyT under contracts MAT2000-1384 and MAT2001-0539.

\end{acknowledgments}



\onecolumn
\begin{table}[tbp]
\caption{Observed and estimated R modes frequencies for RMnO$_{3}$ compounds. The
estimated values have been obtained with $\omega _{R}$=$\omega _{Y}\sqrt{M_{Y}/M_{R}}$. TW 
stands for "this work". (*) this peak is assigned in Ref. 16
 to a R mode (A$_{g}$ symmetry)}
\label{tabla1}
\begin{tabular}{c|ccc|cc|cc|cc|cc|cc|cc}
\multicolumn{1}{c|}{}& \multicolumn{3}{c|}{Ca (40)} & \multicolumn{2}{c|}{Y (89)} & 
\multicolumn{2}{c|}{La (139)} & \multicolumn{2}{c|}{Nd (144)} & 
\multicolumn{2}{c|}{Tb (159)} & \multicolumn{2}{c|}{Ho (165)} & 
\multicolumn{2}{c}{Er (167)} \\ 
assignment&Est.&TW&Ref. 15&TW&Ref. 11&Est.&TW&Est.&TW&Est.&TW&Est.&TW&Est.&TW\\ \hline
A$_{g}$\par/B$_{1g}$&220 & - & 242\par/259 & 148 & 151\par/151 & 118 & 109 & 116.5 & - & 110.5
& 111 & 108.8 & 109 & 108 & - \\ 
A$_{g}$&274 & - & 278 & 184 & 188 & 147 & 141 & 144.9 & 143 & 137 & 138 & 135.2 & 136
& 134 & 134 \\ 
B$_{1g}$&325 & 322 & 322$^{*}$ & 218 & 220 & 174 &  & 171.6 & - & 162.5 & 160 & 160.3 & 160 & 
159.1 & 157
\end{tabular}
\end{table}


\begin{references}
\bibitem{1}  S. Jin, T.H. Tiefel, M. McCormack, R.A. Fastnacht, R. Ramesh,
L.H. Chen, Science 264, 413 (1994)

\bibitem{ziese}  M. Ziese, Rep. Prog. Phys. 65, 143 (2002)

\bibitem{dagotto}  E. Dagotto, T. Hotta and A. Moreo, Phys. Reports 344, 1
(2001)

\bibitem{jtph}  Philip B. Allen and Vasili Perebeinos, Phis. Rev. B 60,
10747 (1999)

\bibitem{5}  J.C. Irwin, J. Chrzanowski and J.P. Franck, Phys. Rev. B 59,
9362 (1999)

\bibitem{6}  E. Liarokapis, Th. Leventouri, D. Lampakis, D. Palles, J.J.
Neumeier and D.H. Goodwin, Phys. Rev. B 60, 12758 (1999)

\bibitem{granado}  E. Granado, A. Garc\'{\i}a, J.A. Sanjurjo, C. Rettori, I.
Torriani, F. Prado, R.D. S\'{a}nchez, A. Caneiro and S.B. Oseroff, Phys.
Rev. B 60, 11879 (1999)

\bibitem{2}  S. Yoon, H.L. Liu, G. Schollerer, S.L. Cooper, P.D. Han, D.A.
Payne, S.W. Cheong and Z. Fisk, Phys. Rev. B 58, 2795 (1998)

\bibitem{3}  H.L. Liu, S. Yoon, S.L. Cooper, S.W. Cheong, P.D. Han and D.A.
Payne, Phys. Rev. B 58, 10115 (1998)

\bibitem{4}  M.R. Ibarra, Guo-meng Zhao, J.M De Teresa, B.
Garc\'{\i}a-Landa, Z. Arnold, C. Marquina, P.A. Algarabel, H. Keller and C.
Ritter, Phys. Rev. B 57, 7446 (1998)

\bibitem{7}  M.N. Iliev, M.V. Abrashev, H.G. Lee, V.N. Popov, Y.Y. Sun, C.
Thomsen, R.L. Meng and C.W. Chu, Phys. Rev. B 57, 2872 (1998)

\bibitem{8}  V.B. Podobedov, A. Weber, D.B. Romero, J.P. Rice, H.D. Drew,
Phys. Rev. B 58, 43 (1998)

\bibitem{RMO} L. Mart\'{\i}n-Carr\'{o}n, A. de Andr\'{e}s, M.T.Casais, M.J.Mart\'{\i}nez-Lope 
and J.A.Alonso, J. Alloys and Compounds 323-324, 494 (2001)

\bibitem{14}  L. Mart\'{\i}n-Carr\'{o}n and A. de Andr\'{e}s, The Europ.
Phys.J. B 22, 11 (2001); L. Mart\'{\i}n-Carr\'{o}n and A. de Andr\'{e}s, J. Alloys and Compounds
323-324, 417 (2001)

\bibitem{amelichev}  V. A. Amelichev, B. G\"{u}ttler, O. Yu. Gorbenko, A. R.
Kaul, A. A. Bosak and A. Yu. Ganin, Phis. Rev. B, 63, 104430 (2001)

\bibitem{ramanca}  M. V. Abrashev, J. B\"{a}ckstr\"{o}m, L. B\"{o}rjesson,
V. N. Popov, R. A. Chakalov, N. Kolev, R.-L. Meng and M. N. Iliev, Phys.
Rev. B, 65, 184301 (2002)

\bibitem{9}  P. Bj\"{o}rnsson, M. R\"{u}bhausen, J. B\"{a}ckstr\"{o}m, M.
K\"{a}ll, S. Eriksson, J. Eriksen, L. B\"{o}rjesson, Phys. Rev. B 61, 1193
(2000)

\bibitem{10}  J.A. Alonso, M.J. Mart\'{\i}nez-Lope, M.T. Casais and M.T.
Fern\'{a}ndez-D\'{\i}az, Inorg. Chem. 39, 917 (2000)

\bibitem{11}  J. Rodr\'{\i}guez-Carvajal, M. Hennion, F. Moussa, A.H.
Moudden, L.Pinsard and A.Revcolevschi, Phys. Rev. B 57 R3189, (1998)

\bibitem{12}  J.A. Alonso, M.J. Mart\'{\i}nez-Lope, M.T. Casais, J.L.
MacManus-Driscoll, P.S.I.P.N. de Silva, L.F. Cohen and M.T.
Fern\'{a}ndez-D\'{\i}az, J. Mater. Chem. 7, 2139 (1997)

\bibitem{13}  J.A. Alonso, Phil. Trans. R. Soc. Lond. A 356, 1617 (1998)

\bibitem{lasr}  A. Urushibara, Y. Moritomo, T. Arima, A. Asamitsu, G. Kido
and Y. Tokura, Phys. Rev. B 51, 14103 (1995)

\bibitem{castr}  K. R. Poeppelmeier, M. E. Leonowicz, J. C. Scanlon, J.M.
Longo and W. B. Yelon, J. Solid State Chem. 45, 71 (1982)

\bibitem{ilievrh}  M.V. Abrashev, A.P. Litvinchuk, M.N. Iliev, R.L. Meng,
V.N. Popov, V.G. Ivanov, R.A. Chakalov and C. Thomsen, Phys. Rev. B 59, 4146
(1999)

\bibitem{deandres}  A. de Andr\'{e}s, S. Taboada, J.L. Mart\'{i}nez, A.
Salinas, J. and R. S\'{a}ez-Puche. Phys. Rev. B 47, 14898-14904 (1993).

\bibitem{paramestr}  P. Dai, J. Zhang, H.A. Mook, S-H. Liou, P.A. Dowben and
E.W. Plummer, Phys. Rev. B 54, R3694 (1996)

\bibitem{paramestr2}  P.G. Radaelli, D.E. Cox, M. Marezio and S-W. Cheong,
Phys. Rev. B 55, 3015 (1997)

\bibitem{paramestr3}  S. Faaland, K.D. Knudsen, M.A. Einarsrud, L. R{\o}
rmark, R. H{\o}ier and T. Grande, J. Solid State Chem. 140, 320 (1998)

\bibitem{16}  S. Shimomura, N. Wakabayashi, H. Kuwahara and Y. Tokura, Phys.
Rev. Lett. 83, 4389 (1999)

\bibitem{louca}  D. Louca and T. Egami, Phys. Rev. B 59, 6193 (1999)

\bibitem{17}  M. Garc\'{\i}a-Hern\'{a}dez, A. Mellerg$\dot{a}$rd, F.J.
Mompean, D. S\'{a}nchez, A. de Andr\'{e}s, R.L. McGreevy and J.L.
Mart\'{\i}nez, cond-mat/0201436

\bibitem{19}  J.M De Teresa, M.R. Ibarra, P.A. Algarabel, C. Ritter, C.
Marquina, J. Blasco, J. Garc\'{i}a, A. del Moral and Z. Arnold, Nature
(London) 386, 256 (1997)

\bibitem{18}  E. Granado, N.O. Moreno, A. Garc\'{\i}a, J.A. Sanjurjo, C.
Rettori, I. Torriani, S.B. Oseroff, J.J. Neumeier, K.J. McClellan, S.W.
Cheong, Y. Tokura, Phys. Rev. B 58, 11435 (1998)
\end{references}
\end{document}